\documentclass[prd,showpacs,showkeys,twocolumn]{revtex4-1}
\usepackage{bm}
\usepackage{amssymb,amsmath,amsthm}
\usepackage{mathrsfs}
\usepackage[applemac]{inputenc}
\begin{document}
%------------------------------------------------
\title{Regge--Wheeler equation, linear stability,  and greybody factors for dirty black holes}
%------------------------------------------------
\author{Petarpa Boonserm}
\email{petarpa.boonserm@gmail.com}
\affiliation{Department of Mathematics and Computer Science, Chulalongkorn University, Bangkok 10330, Thailand}
%------------------------------------------------
\author{Tritos Ngampitipan}
\email{tritos.ngampitipan@gmail.com}
\affiliation{Department of Physics, Chulalongkorn University, Bangkok 10330, Thailand}
%-----------------------------------------------
\author{Matt Visser}
\email{matt.visser@msor.vuw.ac.nz}
\affiliation{School of Mathematics, Statistics, and Operations Research, 
Victoria University of Wellington, PO Box 600, Wellington 6140, New Zealand}
%------------------------------------------------
%\author{\ } %to get some semi-decent spacing
%------------------------------------------------
\date{7 May 2013;  26 July 2013; \LaTeX-ed  \today}
%-----------------------------------------------
\begin{abstract}
%------------------------------------------------
So-called “dirty” black holes are those surrounded by non-zero stress-energy, rather than vacuum. The presence of the non-zero stress-energy modifies key features of the black hole, such as the surface gravity, Regge--Wheeler equation, linear stability, and greybody factors in a rather nontrivial way.  Working within the inverse-Cowling approximation, (effectively the test-field limit), we shall present general forms for the Regge--Wheeler equation for linearized spin~0, spin~1, and axial spin~2 perturbations on an arbitrary static spherically symmetric background spacetime. Using very general features of the background spacetime, (in particular the classical energy conditions for the stress-energy surrounding the black hole), we extract several interesting and robust bounds on the behaviour of such systems, including rigorous bounds on the greybody factors for dirty black holes.

%------------------------------------------------
\end{abstract}
%------------------------------------------------

%-----------------------------------------------------------------
%-----------------------------------------------------------------
\keywords{Dirty black holes; Regge--Wheeler equation; energy conditions; greybody factors\\
Physical  Review D {\bf88} (2013) 041502(R);  doi: 10.1103/PhysRevD.88.041502}
%-----------------------------------------------------------------

%------------------------------------------------
%------------------------------------------------
\maketitle
%------------------------------------------------
%------------------------------------------------

%------------------------------------------------------------------------------------------------------------------------------------------
% very standard definitions
%------------------------------------------------------------------------------------------------------------------------------------------
\newcommand{\scri}{\mathscr{I}}
\newcommand{\sun}{\ensuremath{\odot}}
\def\d{{\mathrm{d}}}
\def\tr{{\mathrm{tr}}}
\def\sech{{\mathrm{sech}}}
\def\etc{\emph{etc}}
\def\ie{{\emph{i.e.}}}
\def\implies{\Rightarrow}
\def\Nordstrom{Nordstr\"om}
\def\I{{\mathcal{I}}}
%------------------------------------------------------------------------------------------------------------------------------------------
%------------------------------------------------------------------------------------------------------------------------------------------
%-----------------------------------------------------------------------------
\def\lint{\hbox{\Large $\displaystyle\int$}} %needs \usepackage{amssymb}
\def\hint{\hbox{\Huge $\displaystyle\int$}}  %needs \usepackage{amssymb}
%-----------------------------------------------------------------------------
\def\Re{{\mathrm{Re}}}
\def\Im{{\mathrm{Im}}}
%------------------------------------------------
%------------------------------------------------
\section{Introduction}
%------------------------------------------------
The “cleanest” black holes to work with are undoubtedly the Schwarzschild and Reissner--\Nordstrom{} black holes. 
However, real physical black holes are typically surrounded by matter or fields of various types, and so are embedded in an environment of non-zero stress-energy. A good model for such systems is a generic static spherically symmetric spacetime with a Killing horizon. These are the so-called “dirty” black holes~\cite{dirty1, dirty2, dirty3}. 
Without any loss of generality, the metric can then be put in the form
\begin{equation}
\d s^2 = -e^{-2\phi(r)} \left[1-{2m(r)\over r}\right]\d t^2 + {\d r^2\over1-2m(r)/r} + r^2 \d\Omega^2.
\end{equation}
The Einstein equations imply
\begin{equation}
m' = 4\pi \rho \,r^2; \qquad
\phi' = - {4\pi(\rho+p_r)\,r \over  1-2m(r)/r}.
\end{equation}
We shall assume the existence of a black hole horizon such that $2m(r_H)=r_H$. 
Furthermore, for simplicity we assume asymptotic flatness, so that $m(\infty)$ is finite, and we can choose $\phi(\infty)=0$. (Asymptotically de~Sitter spacetimes have an additional cosmological horizon $2m(r_C) = r_C$, where we can choose $\phi(r_C)=0$; asymptotically anti-de~Sitter spacetimes exhibit extra technical complications.)

For an asymptotically flat dirty black hole the surface gravity can easily be extracted from a straightforward calculation~\cite{dirty1}:
\begin{equation}
\kappa =  {e^{-\phi(r_H)}\over 2 r_H} [1-2m'(r_H)]. 
\end{equation}
We shall now seek to say as much as we can about these dirty black holes, without making any particular commitment as to the specific equation of state or other but the most general features of the surrounding matter.

%------------------------------------------------
\section{Classical energy conditions}
%------------------------------------------------
While the classical energy conditions are now known to not be fundamental physics~\cite{twilight}, (they are typically violated by semiclassical quantum effects~\cite{Visser:1994, gvp1, gvp2, gvp3, gvp4, gvp5, Flanagan:1996}), they are nevertheless a good first approximation when dealing with bulk matter and/or classical field configurations.  In particular for the weak and null energy conditions we have:
\begin{equation}
\hbox{WEC} \;\;\implies\;\; \rho\geq 0 \quad\implies\quad m(r_H) \leq m(r) \leq m(\infty);
\end{equation}
\begin{equation}
\hbox{NEC} \;\;\implies\;\; \rho+p_r\geq 0 \quad\implies\quad \phi(r_H) \geq \phi(r) \geq 0.
\end{equation}
Note the weak energy condition (WEC) implies the null energy condition (NEC), so the WEC implies that $\kappa \leq 1/(2r_H)$, independent of the specific nature of the matter surrounding the black hole~\cite{dirty1}. It is this sort of model-independent result that we shall now extend first to the Regge--Wheeler equation, and subsequently to explicit bounds on the the greybody factors.

%------------------------------------------------
\section{Regge--Wheeler equation}
%------------------------------------------------
Define a generalized \emph{tortoise coordinate} $r_*$ by
\begin{equation}
{\d r_*\over \d r } = e^{+\phi(r)} \left[1-{2m(r)\over r} \right]^{-1}.
\end{equation}
Then the spacetime metric can be written as
\begin{equation}
\d s^2 = e^{-2\phi(r)} \left[1-{2m(r)\over r}\right] \{-\d t^2 + \d r_*^2\}  + r^2 \d\Omega^2,
\end{equation}
where $r$ is now implicitly viewed as a function of $r_*$. 

%------------------------------------------------
\subsection{Spin zero}
%------------------------------------------------
For a minimally coupled spin zero massless scalar field it is now a simple exercise to show that linearized perturbations are governed by a simple variant of the Regge--Wheeler equation
\begin{equation}
\left[ {\d^2\over \d r_*^2} +  \omega^2 - V(r_*) \right]\psi = 0,
\end{equation}
where now
\begin{equation}
V(r_*) =  e^{-2\phi(r)} \left[1-{2m(r)\over r}\right]{\ell(\ell+1)\over r^2}   + {1\over r} {\d^2 r \over \d r_*^2}.
\end{equation}
If one is considering a scalar field coupled to gravity \emph{with no other matter present} then this result is known to be correct with the provision that $\phi(r)$ and $m(r)$ be set to values consistent with a background solution of the coupled gravity-scalar equations; which in view of the ``no hair'' theorems implies the background is Schwarzschild.  When other non-trivial matter is present the result quoted above holds only within a variant of the inverse-Cowling approximation (wherein fluctuations of the matter fields and spacetime geometry are assumed negligible  compared to fluctuations in the scalar field of interest; see Samuelsson and Andersson~\cite{Samuelsson:2006} for relevant discussion.) This can alternatively be rephrased as saying that we are considering linearized scalar perturbations in the test-field limit.

Application of the Einstein equations (to the background geometry) now yields
\begin{equation}
\label{E:d2r}
{1\over r} {\d^2 r \over \d r_*^2} =  e^{-2\phi(r)} \left[1-{2m(r)\over r}\right]\; \left[{2m(r)\over r^3}  - 4\pi(\rho-p_r)\right],
\end{equation}
whence
 \begin{eqnarray}
V(r_*) &=&  e^{-2\phi(r)} \left[1-{2m(r)\over r}\right]\nonumber\\
&&\times \left[{\ell(\ell+1)\over r^2} + {2m(r)\over r^3}  - 4\pi(\rho-p_r)\right].
\end{eqnarray}
This is clearly consistent with, and a significant generalization of, the standard Schwarzschild result.

%------------------------------------------------
\subsection{Spin one}
%------------------------------------------------
For the spin one Maxwell field a straightforward calculation yields
\begin{equation}
V(r_*) =  e^{-2\phi(r)} \left[1-{2m(r)\over r}\right]{\ell(\ell+1)\over r^2}.
\end{equation}
The correctness of this result may easily be verified \emph{a posteriori} by noting that, due to the conformal invariance of the Maxwell equations in 3+1 dimensions, the physics can depend only on the ratio $e^{-2\phi}(1-2m/r)/r^2$.  Comparison with the known Schwarzschild result then fixes the proportionality constant. 

If one is considering a Maxwell field coupled to gravity \emph{with no other matter present} then this result is known to be correct with the provision that $\phi(r)$ and $m(r)$ be set to values consistent with a background solution of the coupled Einstein--Maxwell equations; which in view of the ``no hair'' theorems implies the background is Reissner--\Nordstrom.  When other non-trivial matter is present the result quoted above holds only within a variant of the inverse-Cowling approximation (wherein fluctuations of the matter fields and spacetime geometry are assumed negligible compared to fluctuations in the Maxwell field). This can be rephrased as saying that we are considering linearized Maxwell perturbations in the test-field limit.

%------------------------------------------------
\subsection{Spin two axial}
%------------------------------------------------
For the case of spin two axial perturbations the calculation is somewhat tedious.   For perfect fluid stars (rather than black holes) there is general agreement that~\cite{Ferrari:2003, Ferrari:2011, Kokkotas:1999}
 \begin{eqnarray}
 \label{E:S2f}
V(r_*) &=&  e^{-2\phi(r)} \left[1-{2m(r)\over r}\right]\nonumber\\
&&\times \left[{\ell(\ell+1)\over r^2} - {6m(r)\over r^3}  + 4\pi(\rho-p)\right].\quad
\end{eqnarray}
Here $p$ is the isotropic pressure; $p=p_r=p_t$ for perfect fluids. For the specific case of boson stars, (with their intrinsically anisotropic stresses), there is a very similar result involving the radial pressure $p_r$~\cite{Kojima:1991}:
\begin{eqnarray}
 \label{E:S2}
V(r_*) &=&  e^{-2\phi(r)} \left[1-{2m(r)\over r}\right]\nonumber\\
&&\times \left[{\ell(\ell+1)\over r^2} - {6m(r)\over r^3}  + 4\pi(\rho-p_r)\right].\quad
\end{eqnarray}
Furthermore, for generic stars supported by anisotropic stress, and subject to the inverse Cowling approximation, (wherein fluctuations of the matter fields are assumed negligible  compared to fluctuations in the spacetime geometry), Samuelsson and Andersson have argued that the above potential (\ref{E:S2}) retains its validity~\cite{Samuelsson:2006}.  

Note that applying the Einstein equations to the background geometry we can rewrite (\ref{E:S2}) as  
\begin{eqnarray}
V(r_*) &=&  e^{-2\phi(r)} \left[1-{2m(r)\over r}\right]  \left[{\ell(\ell+1)\over r^2} - {4m(r)\over r^3} \right]
\nonumber\\
&& - {1\over r} {\d^2 r \over \d r_*^2}.
\end{eqnarray}
Formally there is no obstruction to now applying this result to other situations such as wormholes or dirty black holes. (The traversable wormhole calculations of S-W~Kim~\cite{Kim:2004, Kim:2008, Kim:2010} likewise implicitly apply a version of the inverse Cowling approximation, and provide another consistency check on the above.)

%------------------------------------------------
\subsection{Spins zero, one, and two}
%------------------------------------------------
Now collecting all these results, we can for $S\in\{0,1,2\}$ write the Regge--Wheeler potential in a unified form  as: 
\begin{eqnarray}
V(r_*)\! &=&\!  e^{-2\phi(r)} \! \left[1-{2m(r)\over r}\right]\!\!\left[{\ell(\ell+1)\over r^2} - {S(S-1) 2m(r)\over r^3} \right]
\nonumber\\
&& + {1-S\over r} {\d^2 r \over \d r_*^2}.
\end{eqnarray}
Equivalently:
\begin{eqnarray}
V(r_*)\! &=&\!  e^{-2\phi(r)} \left[1-{2m(r)\over r}\right]  \Bigg[{\ell(\ell+1)\over r^2} + {(1-S^2) 2m(r)\over r^3} 
\nonumber\\
&& - (1-S)4\pi(\rho-p_r) \Bigg].
\end{eqnarray}
We now have a very general version of the Regge--Wheeler potential simultaneously applicable (within the inverse Cowling approximation) to minimally coupled massless scalars, Maxwell fields, and axial perturbations of the spacetime geometry --- for arbitrary static spherically symmetric spacetimes --- and so in particular applicable to (static spherically symmetric) dirty black holes. 

%------------------------------------------------
\section{Stability considerations}
%------------------------------------------------
It is well known that spacetime is linearly stable against oscillations of this type (working within the inverse Cowling approximation) if and only if the Regge--Wheeler equation has no “negative energy” bound states,  (which would correspond to pure imaginary eigenfrequencies).  A \emph{sufficient} condition for stability is $V(r_*) \geq 0$. (Thus stability is automatic for $S=1$, and will need a little further thought for $S=0$ and $S=2$.)  Furthermore, in view of Simon’s theorem on the existence of bound states~\cite{Simon}, a \emph{necessary} condition for stability is $\int_{-\infty}^{+\infty} V(r_*) \d r_* \geq 0$. This same integral also appears in a rather different context --- it controls one of the very general and simple lower bounds one can place on the greybody factors~\cite{greybody}. For this reason we will merge the stability discussion with that below.

%------------------------------------------------
\section{Transmission bounds}
%------------------------------------------------
For one-dimensional potential scattering there are a number of very general and robust bounds that can be placed on the transmission and reflection probabilities~\cite{bounds}. Further developments in generic contexts can be found in~\cite{miller-good, bogoliubov,  analytic, shabat-zakharov}.   For specific applications to black hole greybody factors see~\cite{greybody}, and further developments in~\cite{Tritos:2012, Tritos:2013}.  Among the various bounds one can develop, two particularly simple ones stand out.  Firstly~\cite{bounds, greybody},
\begin{equation}
T(\omega) \geq \sech^2 \left\{ {1\over2\omega} \int_{-\infty}^{+\infty} V(r_*) \d r_* \right\}. 
\end{equation}
Secondly, for any (possibly even rather crude) upper bound on the Regge--Wheeler potential of the form
\begin{equation}
\forall r_* \quad V(r_*) \leq V_* \leq \omega^2,
\end{equation}
we have~\cite{greybody}
\begin{equation}
T(\omega) \geq  1 - {V_*^2\over(2\omega^2-V_*)^2} \geq 1 - {V_*^2\over\omega^4}. 
\end{equation}
The second bound is the more constraining at ultra-high frequencies, while the first bound continues to hold for arbitrarily low frequencies. 

We make no particular claim that these bounds are in any sense optimal, but they are certainly robust,  and make absolutely minimal assumptions regarding the form of the Regge--Wheeler potential (and so implicitly make absolutely minimal assumptions regarding the nature of the stress-energy tensor surrounding the black hole).

%------------------------------------------------
\subsection{Exponential bound}
%------------------------------------------------
Consider the integral 
$ \int_{-\infty}^{+\infty} V(r_*) \,\d r_*$. This can be bounded in the following manner: Observe
\begin{eqnarray}
V(r_*) \d r_* &=&     e^{-\phi(r)} \left[ {\ell(\ell+1)\over r^2}- {S(S-1)2m(r)\over r^3}  \right]  \d r
\nonumber\\
&&\!\! +  {(1-S)\over r}
{\d\over\d r} \left[ e^{-\phi(r)}\left(1-{2m(r)\over r}\right) \right] \d r, \qquad
\end{eqnarray}
which, (temporarily suppressing the argument $r$), equals
\begin{eqnarray}    
&&\hspace{-10pt} 
e^{-\phi} 
\left[ {\ell(\ell+1)\over r^2}- {S(S-1)2m\over r^3} +{(1-S)\over r^2}   \left(1-{2m\over r}\right) \right] \d r 
\nonumber\\
&&
\qquad +  (1-S)
{\d\over\d r} \left[ {1\over r} e^{-\phi}\left(1-{2m\over r}\right) \right] \d r.
\end{eqnarray}
Then, in view of assumed boundary conditions at $r_H$ and at spatial infinity, the total derivative term drops out of the integral so we have (still an exact result) 
\begin{equation}
\int_{r_H}^\infty {e^{-\phi}\over r^2} 
\left[ \ell(\ell+1)+(1-S) - (S-1)^2 \; {2m\over r}  \right] \d r.
\end{equation}
We shall now bound this integral from above and below. 

On the one hand, merely from the definition of horizon, we must have $2m(r)/r < 1$ for $r> r_H$.  Therefore
\begin{equation}
\int_{-\infty}^{+\infty} V(r_*) \,\d r_* > \int_{r_H}^\infty {e^{-\phi(r)}\over r^2} 
\left[ \ell(\ell+1)+S(1-S)  \right] \d r.
\end{equation}
But the relevant multipole indices $\ell$ satisfy $\ell\geq S$, so
\begin{equation}
\int_{-\infty}^{+\infty} V(r_*) \,\d r_* >  2 S \int_{r_H}^\infty {e^{-\phi(r)}\over r^2} 
 \d r \geq 0.
\end{equation}
Thus the \emph{necessary} condition for linearized stability is always satisfied. (Under the stated conditions, without additional assumptions, we cannot \emph{guarantee} linearized instability no matter how weird our matter content is.) 

On the other hand, the WEC implies $\rho\geq 0$, so $2m(r) \geq 2m(r_H) = r_H$, and we see that the integral is bounded above by
\begin{equation}
\int_{r_H}^\infty {e^{-\phi(r)}\over r^2} 
\left[ \ell(\ell+1)+(1-S) - (S-1)^2 {r_H\over r}  \right] \d r.
\end{equation}
But the NEC implies $\phi(r) \geq 0$ and so $e^{-\phi(r)} \leq 1$. Checking that, (because $\ell\geq S$), the integrand remains positive we see the integral is bounded above by
\begin{equation}
\int_{r_H}^\infty {1\over r^2} 
\left[ \ell(\ell+1)+(1-S) - (S-1)^2 {r_H\over r}  \right] \d r.
\end{equation}
But this integral can now be performed explicitly, so
\begin{equation}
\int_{-\infty}^{+\infty} V(r_*) \,\d r_* \leq 
{1\over r_H} \left[ \ell(\ell+1)+(1-S) - {(S-1)^2\over2} \right].
\end{equation}
That is
\begin{equation}
\int_{-\infty}^{+\infty}  V(r_*) \d r_* \leq  
{1\over r_H} \left[ \ell(\ell+1)+ {1-S^2\over2} \right].
\end{equation}
Thence
\begin{equation}
T(\omega) \geq \sech^2\left\{ {1\over 2 \omega\; r_H} \left[ \ell(\ell+1)+ {1-S^2\over2} \right] \right\}.
\end{equation}
We thus have a largely model independent bound on the greybody factor, valid for all frequencies, with minimal assumptions regarding the material external to the black hole. We need to know the black hole radius $r_H$, to know that the exterior matter satisfies the WEC, and know the multipole of interest and spin of the field, but that’s all. 

%------------------------------------------------
\subsection{Polynomial bound}
%------------------------------------------------
For the polynomial bound one needs to place an upper bound on the Regge--Wheeler potential itself, not just its integral. 
For $S=1$ it is elementary that $V(r_*)\geq0$ and (applying the NEC and WEC) that $V(r_*) \leq \ell(\ell+1)/r_H^2$. For $S=0$ and $S=2$ the calculation is less elementary. 
 If we assume the dominant energy condition (DEC) then
\begin{equation}
\hbox{DEC} \quad \implies \quad 0 \leq \rho-p_r \leq 2\rho,
\end{equation}
and consequently (\ref{E:d2r}) implies
\begin{equation}
{1\over r} {\d^2 r \over \d r_*^2} \leq  e^{-2\phi(r)} \left[1-{2m(r)\over r}\right] {2m\over r^3} \leq {1\over r_H^2}.
\end{equation}
More subtly
\begin{equation}
{1\over r} {\d^2 r \over \d r_*^2} \geq  - e^{-2\phi(r)} \left[1-{2m(r)\over r}\right] {2\over r} \left( {m(r)\over r}\right)'.
\end{equation}
If we now make the \emph{additional} assumption that one has $\left( {m(r)/ r}\right)' \leq 0$,   (which is not unreasonable but certainly nontrivial), then 
\begin{equation}
0 \leq {1\over r} {\d^2 r \over \d r_*^2} \leq {1\over r_H^2}.
\end{equation}
Then for $S=0$
\begin{equation}
0 \leq V(r_*) \leq {\ell(\ell+1)+1\over r_H^2},
\end{equation}
while for $S=2$ 
\begin{equation}
0 \leq V(r_*) \leq {\ell(\ell+1)-2\over r_H^2}.
\end{equation}
Collecting these three results for $S\in\{0, 1, 2\}$, we have
\begin{equation}
0 \leq V(r_*) \leq  {\ell(\ell+1) + {1\over2}(1-S)(2+S) \over r_H^2}.
\end{equation}
So under these conditions we are guaranteed linearized stability within the inverse Cowling approximation, \emph{and}
we have the explicit bound
\begin{equation}
T(\omega) \geq 1 - {\{\ell(\ell+1) + {1\over2} (1-S)(2+S)\}^2 \over r_H^4\; \omega^4}. 
\end{equation}
Again we have a very general and robust bound based on minimal input assumptions. 

\enlargethispage{40pt}
%------------------------------------------------
\section{Discussion}
%------------------------------------------------
As always with generic results there is a trade-off between generality and specificity. In this article we have attempted to be as general as possible, using only relatively weak constraints on the spacetime geometry to still extract very general and useful information regarding linear stability and the greybody factors. 
Of course, any explicit choice for the functions  $\phi(r)$ and $m(r)$ will, at least in principle, allow one to extract much more specific results. 
Additionally it is conceivable that the general techniques of \cite{bounds, greybody, miller-good, bogoliubov, analytic, shabat-zakharov, Tritos:2012, Tritos:2013} could be further extended to obtain more stringent bounds.

Furthermore, the results of this article can be viewed as placing bounds on the behaviour of the Regge--Wheeler operator (which would generically be part of any perturbation scheme that seeks to go beyond the inverse Cowling approximation).
A specific choice of matter model would in certain situations allow one to move beyond the inverse Cowling approximation, but at the cost of massive complications due to possible couplings between various perturbative sectors, and with consequent massive loss of generality.  
Finally, an extension to spin two polar perturbations described by a generalized Zerilli-type equation is in principle certainly possible (see for instance~\cite{Kim:2010}), but is mathematically somewhat messier. 

\newpage
%------------------------------------------------
\section*{Acknowledgments}
%------------------------------------------------

PB was supported by a grant for the professional development of new academic staff from 
the Ratchadapisek Somphot Fund at Chulalongkorn University, by the Thailand Toray Science Foundation (TTSF), by the Thailand Research Fund (TRF), and by the Research Strategic plan program (A1B1), Faculty of Science, Chulalong\-korn University. TN was supported by a scholarship from the Development and Promotion of Science and Technology talent project (DPST). MV was supported by the Marsden Fund, and by a James Cook fellowship, both administered by the Royal Society of New Zealand.

%----------------------------------------------------------------------
%----------------------------------------------------------------------

%-----------------------------------------------------------------------

%------------------------------------------------
\end{document}